# A trapped field of 17.7 T in a stack of high temperature superconducting tape

**Anup Patel[1], Algirdas Baskys[1], Tom Mitchell-Williams[1], Aoife McCaul[1], William Coniglio[2], Bartek A Glowacki[1,3]**

[1]Applied Superconductivity and Cryoscience Group, Department of Materials Science and Metallurgy, University of Cambridge, 27 Charles Babbage Road, Cambridge, CB3 0FS, UK

[2]National High Magnetic Field Laboratory, Tallahassee, Florida 32310, USA

[3]Institute of Power Engineering, ul. Mory 8, 01-330 Warsaw, Poland

Email: ap604@cam.ac.uk

**Abstract:**

High temperature superconducting (HTS) tape can be cut and stacked to generate large magnetic fields at cryogenic temperatures after inducing persistent currents in the superconducting layers. A field of 17.7 T was trapped between two stacks of HTS tape at 8 K with no external mechanical reinforcement. 17.6 T could be sustained when warming the stack up to 14 K. A new type of hybrid stack was used consisting of a 12 mm square insert stack embedded inside a larger 34.4 mm diameter stack made from different tape. The magnetic field generated is the largest for any trapped field magnet reported and 30% greater than previously achieved in a stack of HTS tapes. Such stacks are being considered for superconducting motors as rotor field poles where the cryogenic penalty is justified by the increased power to weight ratio. The sample reported can be considered the strongest permanent magnet ever created.

## 1. Introduction

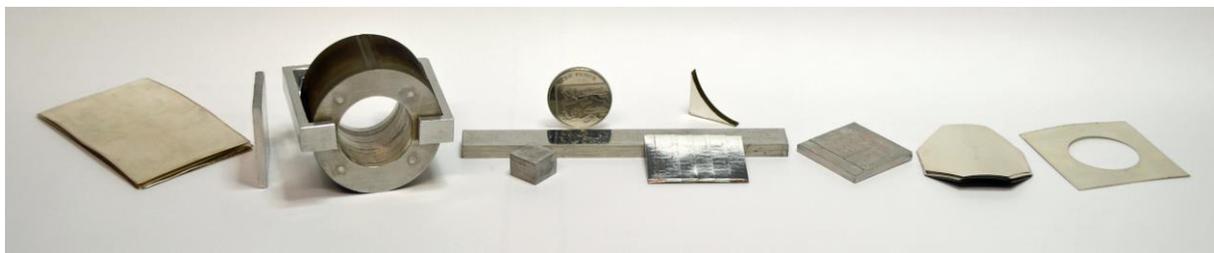

**Figure 1**: Selection of various stacks of HTS tape and tape pieces illustrating the range of shapes and sizes of composite bulk that can now be produced.

HTS tape produced commercially can sustain large persistent currents in their micron thick superconducting layers which have the chemical composition, $(RE)Ba_2Cu_3O_{7-\delta}$ or simply (RE)BCO, where RE is a rare earth metal, most commonly Y or Gd or a combination of both. Although primarily manufactured by a number of global suppliers for use in cables and coils, when cut into pieces and stacked, composite bulks can be formed as illustrated in Figure 1, which can trap high magnetic fields despite only containing a few percent HTS by volume. Most previous trapped field magnets have taken the form of bulk (RE)BCO [1, 2], grown via top seeded melt growth, where the volume fraction of HTS is typically > 90%. This means that very high engineering current densities are possible, which has allowed trapped fields of up to 17.6 T at 26 K [1], the previous record field for a trapped field magnet. However, their limiting property is mechanical strength due to cracks in the brittle ceramic HTS. These arise due to the large tensile stresses inside the sample resulting from Lorentz forces. External mechanical reinforcement is essential to trap fields > 17 T in a bulk, and even then it is common for a number of samples in a batch to break when trapping such high fields.

The large metallic volume fraction and layered structure of stacks of HTS tape have numerous advantages for trapping very high fields and for applications where such composite bulks could be used as permanent magnets. The four main advantages are as follows. i) Their geometry is very flexible and they can be easily machined allowing many shapes. ii) The superalloy substrates, which account for more than 85 % of the volume fraction, have a very high tensile strength, which means no external mechanical reinforcement is needed to counter Lorentz forces at high trapped fields. iii) The superconducting properties are generally consistent throughout the volume of the stack and defects in individual layers are smoothed out in the trapped field profiles, meaning that different stacks made from the same batch of tape have the same performance. iv) The silver stabilizer layer on top of the HTS layer provides thermal stability which helps dissipate heat generated inside the stack and supresses flux jumps.

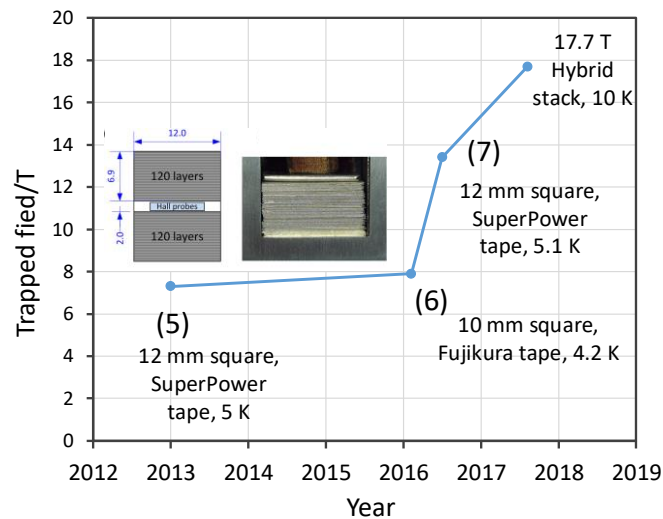

**Figure 2**: Trapped field records for stacks of HTS tape. Field cooling magnetization used and field measured between two stacks according to convention.

Initial experiments on trapping field in stacks of HTS tape were performed in liquid nitrogen at 77.4 K [3]. However, because the critical current of the HTS layer increases significantly as you lower the temperature below 77 K, further trapped field experiments were performed at lower temperatures [4-7]. Progress in the largest field trapped between two stacks of tape is shown in Figure **2**, demonstrating the rapid progress in the maximum trapped field achieved. As the critical current of commercially produced tapes steadily improves, further progress is inevitable.

## 2. Stack composition and magnetization procedure

### 2.1. Stack fabrication and composition

The magnetic field generated by a trapped field magnet is fundamentally limited by two factors due to Ampere's law; the diameter of the magnet (assuming the height is unconstrained) and the engineering current density $J_e$ (A/m$^2$). The previous trapped field records for stacks of tape [4, 6, 7] were all achieved using 12 mm or 10 mm square stacks as this is the standard width of tape most suppliers produce. The thin 30 µm substrate used by SuperPower Inc. has allowed for significant increases in $J_e$. Only American Superconductor (AMSC) currently produces wider tape routinely, which is then typically slit down to standard 12 or 4 mm wide HTS wire. To combine the advantages of the highest $J_e$ stacks with a larger width stack, a hybrid design was used. A high $J_e$ stack made from SuperPower tape was embedded inside a lower $J_e$, but larger size stack made from 46 mm wide AMSC tape. The geometry and composition of the hybrid stack is shown in Figure 3.

The SuperPower stack was made from tape with specification SP12030 AP and composition $(Y,Gd)_{1+x}Ba_2Cu_3O_{7-\delta}$ with 7.5% Zr added. The rated critical current, $I_c$, was 540 A for 12 mm width (77 K, self-field) corresponding to a $J_e$ of $1.32 \times 10^9$ A/m$^2$. The AMSC tape had an $I_c$ of 391 A per centimetre width at (77 K, self-field) which corresponds to a $J_e$ of $4.49 \times 10^8$ A/m$^3$, almost 3 times lower largely due to the much thicker substrate. The tensile yield strengths of the tape substrates determine the trapped field mechanical limit. HTS tape can usually go beyond its elastic limit with only a few % degradation in $I_c$ [8], but for the purposes of trapped field magnets we can consider the yield stress as the maximum acceptable stress. The SuperPower tape substrate, Hastelloy C276, has a tensile yield strength of 700 MPa and the AMSC substrate, Ni-5at%W, has a yield strength of 257 MPa [8]. Detailed FEM modelling of the stress in the hybrid stack [9] predicted that it would take a trapped field of 31.5 T to reach the mechanical limits of the stack. The AMSC tape is the limiting mechanical component for the hybrid stack. This shows that mechanical properties are not a limiting factor for the current stacks.

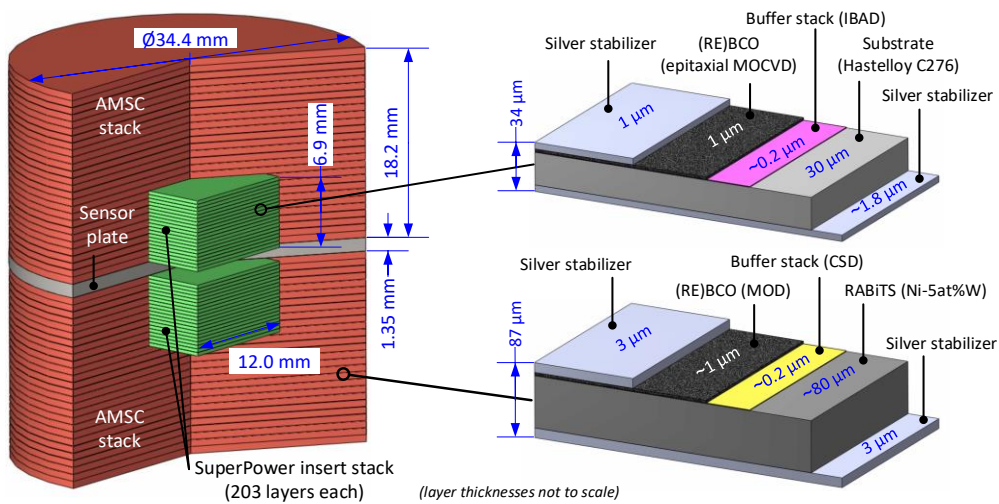

**Figure 3**: Composition and geometry of the hybrid stack consisting of a 12 mm square stack made from SuperPower tape inside a larger cylindrical stack made from American Superconductor tape.

Although the AMSC tape had a width of 46 mm, the maximum external diameter of the stack was limited by the bore of the superconducting magnet used for magnetization. The stack was therefore machined to an external diameter of 34.4 mm, which was the maximum possible size, using the technique of spark erosion as illustrated in Figure 4. This technique has been used previously to machine an AMSC stack for magnetic levitation applications with the cut damage to the HTS layer measured to be less than 0.4 mm from the cut surface [10]. It is a highly precise method for machining stacks that was also used to cut out the square void for the SuperPower stack as shown in Figure 5. The erosion weakly fuses the tape layers together on the cut surface which aids their integrity when handling without external constraints as shown in Figure 5. The SuperPower layers were loosely stacked inside the square cavity before placing the 129 layer AMSC stack on top. Machining the SuperPower stack into a cylinder and inserting it into a cylindrical hole was a possibility that would have enhanced trapped field, as the current circulating in the AMSC stack would not have been constrained by the square corners. However, this would have risked damaging the SuperPower tape, which generates most of the central trapped field. Therefore, this option was not considered worth the risk but could be tested in future. Interestingly, the central trapped field from a 12 mm square sample compared to a 12 mm diameter circular one, are very similar (a few % different at most) based on FEM modelling, so this alone does not affect the central trapped field.

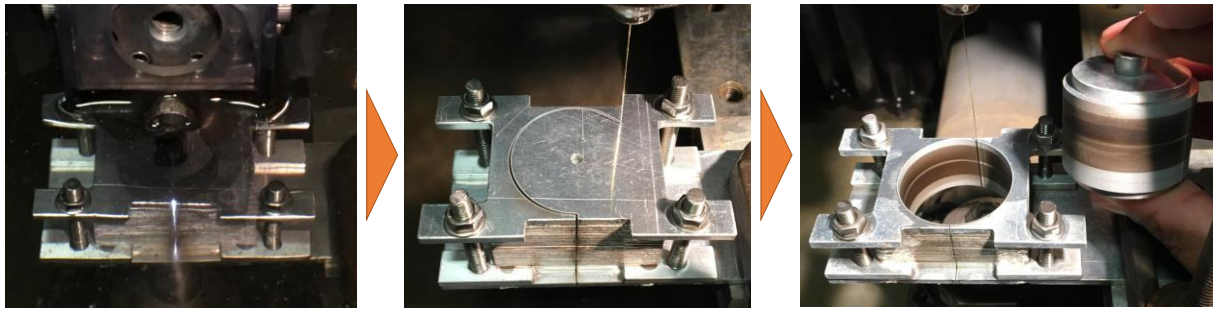

**Figure 4**: Spark erosion machining of parts of the American Superconductor external stack from 46 mm wide tape.

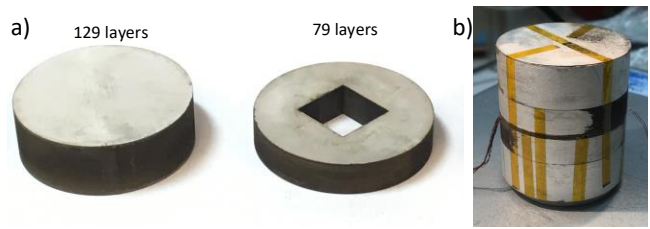

**Figure 5**: a) Components of one half of the American Superconductor stack after spark erosion machining. b) Final hybrid stack fully assembled.

*2.2. Instrumentation and measurement procedure*

A stainless steel sensor plate was placed in between the stacks and contained three Arepoc model LHP-MP cryogenic Hall probes at positions $x = 0$, 6 and 13.5 mm away from the centre. The 6 mm position lies exactly at the edge of the insert stack. Due to the large number of stack layers, the trapped field is expected to be highly symmetric [11] and so it is sufficient to have Hall probes only on one side. These probes are low sensitivity and have very high linearity (< 2% deviation at 18 T), so are specifically suited to trapped field measurement and are reliable after thermal cycling. The cryogenic Hall probes have a relatively large package size (about 5 mm in width) which unfortunately mean more than 3 could not fit between the centre and sample side. Reliability of the field measurement was chosen over resolution of the trapped field profile. A fully calibrated Lakeshore Cernox CX-1070 temperature sensor was also embedded in the sensor plate. The thickness of the plate was chosen to fully carry the large compressive force between the stacks when magnetized to prevent any compression of the sensors. The stacks and sensor plate were held together by insertion in a stainless steel cylinder which applied a gentle axial compression to hold the components together.

The Hall probes were fully calibrated once at 100 K up to 18 T in the superconducting magnet before the magnetization tests, accounting for any non-linearity. They were driven with an AC current of 17.5 mA at 321 Hz using a Stanford Research Systems voltage controlled current source, and the Hall voltages were measured using lock in amplifiers. Field cooling magnetization was used to trap field where the sample is held above its critical temperature (100 K) whilst the magnet is ramped up to the desired applied field. The sample is then cooled to a temperature at which it is superconducting before ramping the field down to zero which induces the persistent currents. The magnetization was achieved using an 18 T low temperature superconducting magnet (SCM-2) at the National High Magnetic Field Laboratory (Florida State University), which is the same magnet used for the previous trapped field record in bulks [1]. Sample temperature was controlled by varying power to a heater that is immersed in a pool of liquid helium. The resulting variable flow of helium vapour determined the temperature of the sample.

## 3. Results and discussion

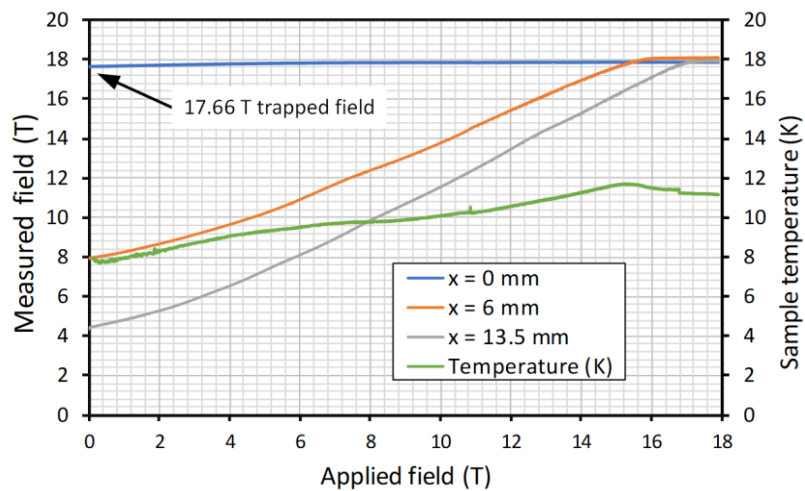

**Figure 6**: The magnetic field measured using the 3 Hall sensors during ramp down of a 17.9 T applied field. The Temperature drifted down to 8 K during the ramp. The trapped fields are the measured fields at the end of the ramp (applied field = 0 T).

The hybrid stack was cooled to 11 K in the presence of a 17.9 T applied field. The field was then ramped down at a rate of 15.5 mT/s. The fields measured by the Hall probes are shown in Figure 6. The applied field choice was based on modelling predictions of the approximate trapped field expected and the trapped field achieved at higher temperatures. Given the relatively low temperatures, flux jumps are a concern and do occur in stacks of HTS tape if too high ramp rates are used as was seen when testing the SuperPower stack alone and in previous tests [7]. Therefore, a conservative ramp rate was used to minimise the risk of flux jumps. The sample temperature proved difficult to completely stabilise below 15 K resulting in a drift, partly because the sample in continuously generating heat during ramp down as flux leaves the sample. Heater settings were chosen to deliberately ensure a negative rather than positive drift to prevent a massive flux avalanche of trapped field due to uncontrolled sample warming. The final sample temperature of 8 K can be considered the temperature at which the field was trapped, with subsequent warming, as reported later, revealing what would have been trapped at temperatures higher than 8 K. The relatively flat line of the central Hall probe indicates that the sample centre is very well shielded from changes in the external applied field which suggests that the sample was either not saturated or only just saturated. Before the ramp down, the off-centre hall probes recorded a field higher than the applied field. This is entirely expected due to the magnetic Ni-W substrate of the AMSC tape which was incorporated into FEM models. These models show that if there is an applied field but no induced currents (the case at the start of the ramp down) then the field across the entire sensor plate surface is enhanced by the ferromagnetic AMSC tape. The trapped field at the centre of the stack ($x = 0$) was measured to be 17.66 T at the end of the ramp.

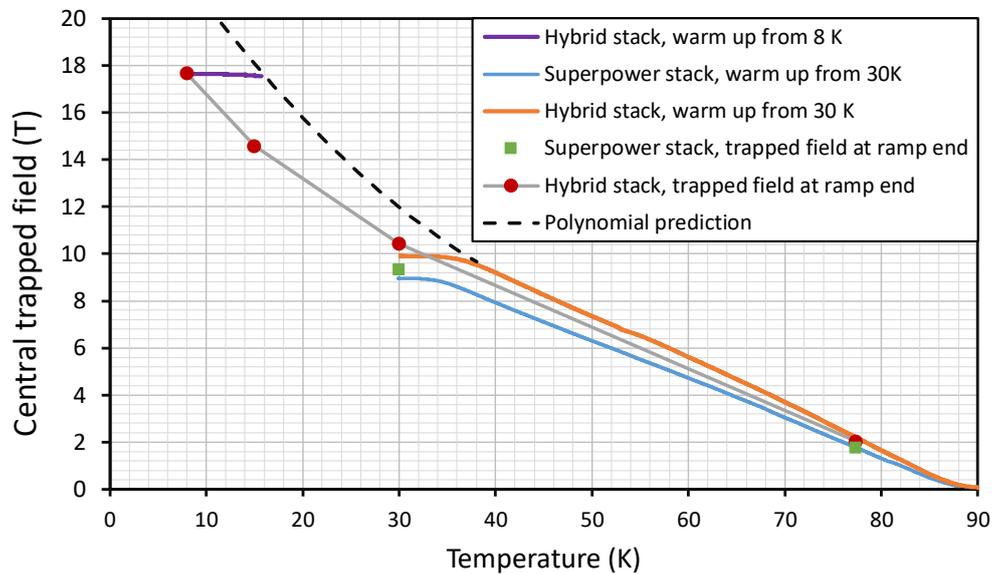

**Figure 7**: Summary of the experimental trapped field results for the hybrid stack and the superpower stack on its own. The solid data points indicate the trapped field at the end of an applied field ramp for a particular temperature. The solid lines are trapped field data as the sample is warmed up from the corresponding temperature. The dashed line is a prediction of the maximum trapped field of the hybrid stack based on a quadratic interpolation of the orange and purple data points.

After the ramp end, the sample was slowly warmed up to 16 K at a rate of 0.1 K/min as shown by the purple data in Figure 7, to determine the maximum trapped field at temperatures slightly higher that the magnetization temperature. Faster warm up rates caused the stack to quench (flux avalanche resulting in a collapse of the trapped field) after the 15 K test. Due to limited magnet time, the warm up from 8 K could not extend all the way to 77 K, however prior tests leading up to the one at 8 K give an overall picture of the trapped field possible at different temperatures as summarised in Figure 7. The trapped fields achieved at 15 K, 30 K and 77.4 K are also shown for which the applied fields were 14.6 T, 13.8 T and 2.5 T respectively with ramp rates of 30, 50 and 300 mT/min used respectively. After trapping a field of 10.4 T at 30 K, the sample was warmed up to 90 K (orange data) showing the maximum field that can be trapped at all higher temperatures. This data set starts at a lower trapped field than the data point for the field at the ramp end. This is due to flux creep at 30 K for 150 minutes after the ramp ended before the warm up was started. Based on all the trapped field data for the hybrid stack, a quadratic interpolation can be made (black dotted line) to give an approximation of the maximum trapped field possible between 10 and 35 K. This interpolation suggests that the hybrid stack was not fully saturated for the 3 tests at 8, 15 and 30 K with a higher trapped field possible if a higher applied field was used. However, based on experience of field cooling experiments, thermal instability and flux jumps are far more common if using a magnetizing field higher than necessary to saturate the sample. This justifies a more cautious approach to maximise informative results given limited magnet time. The overall data for the hybrid stack suggest that a trapped field of more than 18 T is likely possible for the sample below 15 K.

In addition to the hybrid stack, the SuperPower stack was magnetized on its own. Figure 7 shows the trapped field at 30 K and subsequent warm up to 90 K. The trapped field of 9.3 T at the end of the ramp is 90% of that achieved for the hybrid stack, indicating that the insert stack is making the major contribution to the trapped field in the hybrid stack. The 9.3 T value is 13% higher than the interpolated trapped field of a similar previous SuperPower stack made from the same type of 30 µm substrate tape [7]. This is partly because the rated $I_c$ of 540 A is higher than the previous tape and also because the gap between the stacks where the sensor plate sits is smaller (1.35 mm compared to 2.0 mm).

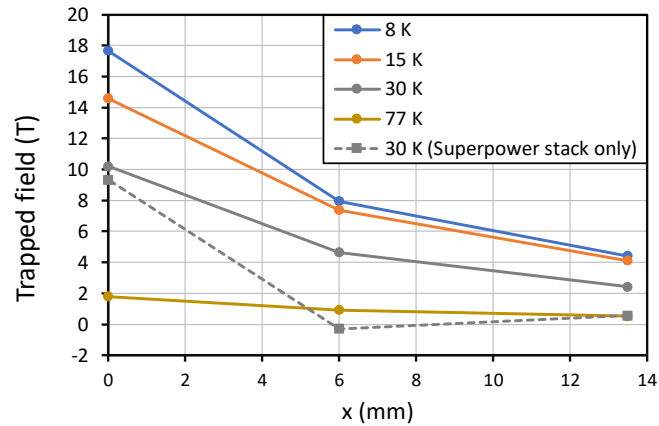

**Figure 8**: Trapped field profiles for the hybrid stack at various magnetization temperatures. The 15 K profile is not saturated. The 30 K result for the SuperPower stack alone is included for comparison, illustrating the majority contribution it makes to the central field.

The trapped field profiles measured by the 3 Hall probes at the end of the ramp are shown in **Figure 8**. Due to the higher engineering current density of the insert stack, the trapped field has a steeper gradient in the central region. The profiles suggest that the trapped field at 15 K is not saturated which is confirmed by the large gap between the trapped field data point for 15 K in **Figure 7**, and the trapped field achieved at the same temperature when warming up the 8 K test (purple data). For comparison, the trapped field profile for the SuperPower stack alone at 30 K is shown in **Figure 8** and indicates that whilst the central trapped field is not much less than that achieved by the hybrid stack, the total trapped flux is much less given the field drops to zero at $x = 6$ mm. This can be more fully visualised by considering the modelled field distribution for both situations shown in Figure 10. This model was constructed in COMSOL Multiphysics 5.3 in the AC/DC module using a stationary study. It can be used to determine the saturated state of a bulk superconductor or stack of tapes expected after field cooling magnetization. A circulating current density in the domains is specified as a function of temperature and field. This function is equal to the effective engineering current density of tapes and gives an approximate prediction of the trapped field. The effect of the non-linear ferromagnetic Ni-W substrate of the AMSC tape was included. The model generally agrees with the experimental results, particularly in showing that the trapped field due to the insert stack alone is not much less than for the hybrid stack. Further details of this model and its use in determining the maximum stresses in the hybrid stack are reported in [9].

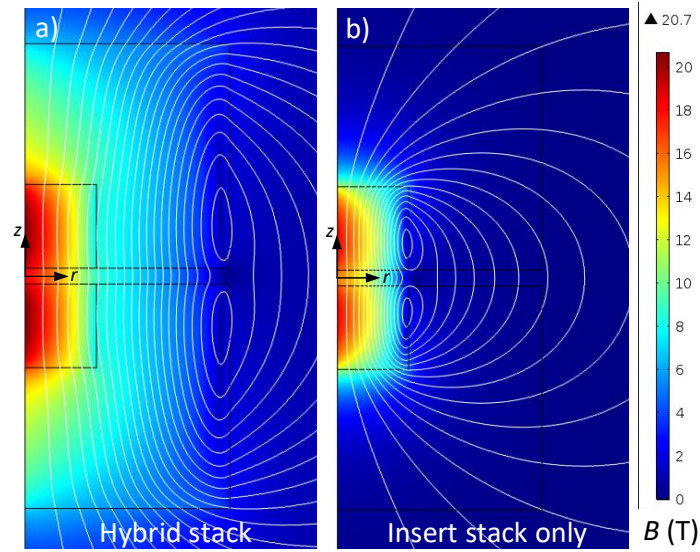

**Figure 9**: Modelling of the critical state of a) the hybrid stack and b) the insert stack only, after field cooling magnetization at 10 K assuming axi-symmetric symmetry.

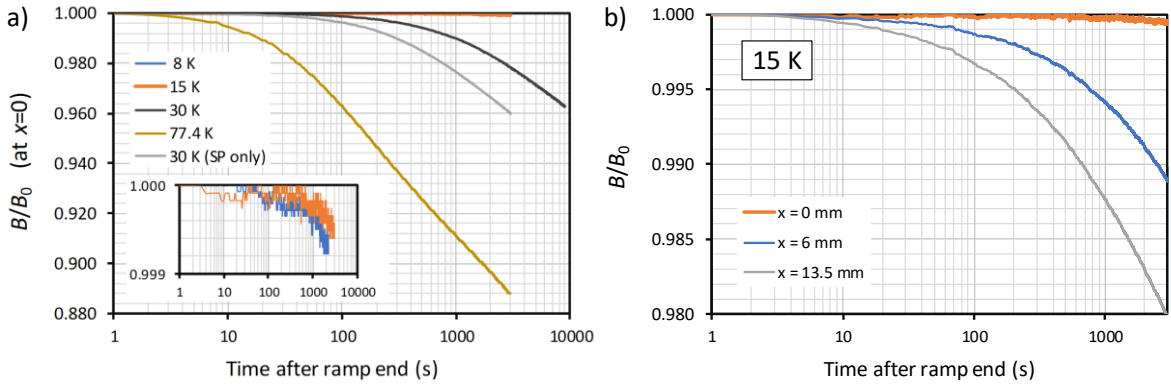

Figure 10: a) Flux creep for the normalised trapped field at the sample centre after magnetization of the hybrid stack at various temperatures. Results for the SuperPower stack only, at 30 K are included for comparison. $B_0$ is the trapped field at $t = 1$s. b) Example flux creep for the normalised trapped field at the 3 different positions for the hybrid stack.

After each magnetization test, the fields were recoded for a period of at least 30 minutes to measure flux creep of the trapped field. Figure 10 shows that the creep rates are lower for lower temperatures which agrees with previous experiments [5, 7] and that the creep is approximately logarithmic after the first ~ 500 s which is expected [12]. The creep of the central trapped field was very low for 15 K and 8 K (less than 0.1% decay after 30 minutes), however the creep of the field at the edges of the sample is higher as illustrated by Figure 10b). This is expected as flux first begins to leave the sample from the outer edges. In general, such flux creep is not a concern for applications of trapped field magnets because the decay in field is logarithmic and also because it can be effectively eliminated by lowering the sample temperature a few Kelvin below the magnetization temperature.

## 4. Summary

A trapped field of 17.7 T was achieved in a hybrid stack of HTS tapes consisting of two different types of tape. This field is slightly higher than the previous record trapped field achieved in a GdBCO bulk [1] making it the strongest trapped field magnet to date. The result is significant given that it is a large increase from the previous trapped field of 13.4 T [7] achieved by a stack of HTS tapes and indicates the rapid progress being made with this type of permanent magnet. The hybrid sample could be quenched and reused without destruction or obvious degradation which is often not the case for bulk superconductors. The sample had relatively predictable trapped field and due to the consistency of HTS tape, another stack made from the same tape can reliably be expected to trap the same field.

Although the insert of the hybrid stack produced most of the central trapped field, the outer AMSC stack still contributed significantly to the total trapped flux based on the trapped field profiles. This makes hybrid stacks valid for motor applications where the total trapped flux is as important as peak trapped field. Improvements in the $I_c$ of HTS tape are expected which is important as the $I_c$ is the main limiting factor for the trapped field rather than mechanical strength. This should allow a similar trapped field to be achieved at temperatures higher than ~ 15 K in future. The practical promise of such stacks of HTS tape is currently being put to the test in the EU project ASuMED: Advanced Superconducting Motor Experimental Demonstrator (Grant No. 723119). This project will test stacks of HTS tape as rotor field poles in a fully superconducting 1 MW motor.

**Acknowledgements**

The authors would like to acknowledge the financial support of EPSRC (grant EP/P000738/1). A portion of this work was performed at the National High Magnetic Field Laboratory, which is supported by National Science Foundation Cooperative Agreement No. DMR-1157490 and the State of Florida.